
\def\ung{{{\frak{g}}}}
\def\ungh{{{{\hat{\frak{g}}}}}}
\def\uqg{{{U_{q}(\ung)}}}
\def\uqgh{{{U_{q}(\ungh)}}}

\def\bp{{{\bold{P}}}}
\def\bq{{{\bold{Q}}}}
\def\calp{{{{\Cal{P}}}}}

\def\ot{{{\otimes}}}
\def\op{{{\oplus}}}

\def\unh{{{\frak{h}}}}

\def\calp{\Cal P}

\def\deg{\rm{deg}}
\input amstex


\def\next{AMSPPT}\ifx\styname\next
  \else\input amsppt.sty\fi

\catcode`\@=11


\pageheight{47pc}
\pagewidth{29pc}
\parindent=12\p@
\def\pretitle{\vskip7pc}



\def\foliofont@{\eightrm}
\def\headlinefont@{\eightpoint}


\def\jourlogo{\vbox to0pt{%
        \sixrm \baselineskip6pt
        \parindent0pt \frenchspacing
        Canadian Mathematical Society\hfil\break
        Conference Proceeding\hfil\break
        Volume {\sixbf\cvol@}, \cvolyear@\par\vss}}


\def\cvol#1{\gdef\cvol@{\ignorespaces#1\unskip}}
\def\cvolyear#1{\gdef\cvolyear@{\ignorespaces#1\unskip}}
\def\cyear#1{\gdef\cyear@{\ignorespaces#1\unskip}\cyear@@#100000\end@}

\def\cyear@@#1#2#3#4#5\end@{\gdef\cyearmodc@{#3#4}%
        \gdef\cyearmodcHold@{#3#4}}

\cvol{00}
\cyear{0000}
\cvolyear{0000}

\font\sixsy=cmsy6

\def\copyrightline@{\baselineskip1.75pc
    \rightline{%
        \vbox{\sixrm \textfont2=\sixsy \baselineskip 7pt
            \halign{\hfil##\cr
                \copyright\cyear@\ American Mathematical Society\cr
                 0000-0000/\cyearmodc@\ \$1.00 + \$.25 per page\cr }}}}

\def\cyearmodc#1{\gdef\cyearmodc@{\ignorespaces#1\unskip}}


\let\logo@=\copyrightline@


\addto\tenpoint{\normalbaselineskip13\p@
 \abovedisplayskip6\p@ plus6\p@ minus0\p@
 \belowdisplayskip6\p@ plus6\p@ minus0\p@
 \abovedisplayshortskip0\p@ plus3\p@ minus0\p@
 \belowdisplayshortskip2\p@ plus3\p@ minus0\p@
 \ifsyntax@
 \else
  \setbox\strutbox\hbox{\vrule height9\p@ depth4\p@ width\z@}%
  \setbox\strutbox@\hbox{\vrule height8\p@ depth3\p@ width\z@}%
 \fi
 \normalbaselines\rm}


\font@\titlebf=cmbx10 scaled \magstep2
\font@\titlei=cmmi10 scaled \magstep2
\font@\titlesy=cmsy10 scaled \magstep2
\def\titlefont{\normalbaselineskip18\p@
 \textonlyfont@\bf\titlebf
 \ifsyntax@\else
  \textfont\z@\titlebf  \scriptfont\z@\tenrm  \scriptscriptfont\z@\sevenrm
  \textfont\@ne\titlei  \scriptfont\@ne\teni  \scriptscriptfont\@ne\seveni
  \textfont\tw@\titlesy \scriptfont\tw@\tensy \scriptscriptfont\tw@\sevensy
  \textfont\thr@@\tenex \scriptfont\thr@@\tenex \scriptscriptfont\thr@@\tenex
 \fi
 \normalbaselines\titlebf}

\def\title{\let\savedef@\title
 \def\title##1\endtitle{\let\title\savedef@\let\\=\cr
   \global\setbox\titlebox@\vtop{\titlefont\bf
   \raggedcenter@\frills@{##1}\endgraf}%
 \ifmonograph@ \edef\next{\the\leftheadtoks}\ifx\next\empty
    \leftheadtext{##1}\fi
 \fi
 \edef\next{\the\rightheadtoks}\ifx\next\empty \rightheadtext{##1}\fi
 }%
 \nofrillscheck\title}


\def\author#1\endauthor{\global\setbox\authorbox@
 \vbox{\tenpoint\raggedcenter@
  \expandafter\uppercase{\ignorespaces#1\endgraf}}\relaxnext@
 \edef\next{\the\leftheadtoks}%
 \ifx\next\empty\expandafter\uppercase{\leftheadtext{#1}}\fi}

\def\address#1\endaddress{\global\advance\addresscount@\@ne
  \expandafter\gdef\csname address\number\addresscount@\endcsname
  {\vskip12\p@ minus6\p@\indent\eightpoint{\smc\ignorespaces#1}\par}}


\def\email{\let\savedef@\email
  \def\email##1\endemail{\let\email\savedef@
  \toks@{\def\usualspace{{\it\enspace}}\endgraf\smallskip\indent\eightpoint}%
  \toks@@{##1\par}%
  \expandafter\xdef\csname email\number\addresscount@\endcsname
  {\the\toks@\frills@{{\noexpand\it E-mail address\noexpand\/}:%
     \noexpand\enspace}\the\toks@@}}%
  \nofrillscheck\email}

\def\curraddr{\let\savedef@\curraddr
  \def\curraddr##1\endcurraddr{\let\curraddr\savedef@
  \toks@\expandafter\expandafter\expandafter{%
       \csname address\number\addresscount@\endcsname}%
  \toks@@{##1}%
  \expandafter\xdef\csname address\number\addresscount@\endcsname
  {\the\toks@\endgraf\noexpand\nobreak
    \smallskip\indent\noexpand\eightpoint{\noexpand\rm
    \frills@{{\noexpand\it Current address\noexpand\/}:\space}%
    \def\noexpand\usualspace{\space}\the\toks@@\unskip}}}%
  \nofrillscheck\curraddr}


\def\abstract{\let\savedef@\abstract\def\abstract{\let\abstract\savedef@
  \setbox\abstractbox@\vbox\bgroup\indenti=3pc\noindent$$\vbox\bgroup
  \def\envir@end{\endabstract}\advance\hsize-2\indenti
  \def\usualspace{\enspace}\eightpoint \noindent
  \frills@{{\smc Abstract.\enspace}}}%
 \nofrillscheck\abstract}


\outer\def\endtopmatter{\add@missing\endabstract
 \edef\next{\the\leftheadtoks}\ifx\next\empty
  \expandafter\leftheadtext\expandafter{\the\rightheadtoks}\fi
 \ifmonograph@\else
   \ifx\thesubjclass@\empty\else \makefootnote@{}{\thesubjclass@}\fi
   \ifx\thekeywords@\empty\else \makefootnote@{}{\thekeywords@}\fi
   \ifx\thethanks@\empty\else \makefootnote@{}{\thethanks@}\fi
 \fi
  \jourlogo
  \pretitle
  \vskip24\p@ plus12\p@ minus12\p@
  \unvbox\titlebox@
  \topskip10pt
  \preauthor
  \ifvoid\authorbox@\else \vskip14\p@ plus6\p@ minus0\p@\unvbox\authorbox@\fi
  \predate
  \ifx\thedate@\empty\else \vskip6\p@ plus2\p@ minus0\p@
    \line{\hfil\thedate@\hfil}\fi
  \preabstract
  \ifvoid\abstractbox@\else \vskip24\p@ plus12\p@ minus0\p@
    \unvbox\abstractbox@ \fi
  \ifvoid\tocbox@\else\vskip1.5pc plus.5pc \unvbox\tocbox@\fi
  \prepaper
  \vskip24\p@ plus12\p@ minus0\p@}


\def\headfont@{\bf}
\def\refheadfont@{\smc}
\let\varindent@\indent


\def\proclaimheadfont@{\smc}

\let\proclaim\relax
\outer\def\proclaim{%
  \let\savedef@\proclaim \let\proclaim\relax
  \add@missing\endroster \add@missing\enddefinition
  \add@missing\endproclaim \envir@stack\endproclaim
 \def\proclaim##1{\restoredef@\proclaim
   \penaltyandskip@{-100}\medskipamount\varindent@
   \def\usualspace{{\proclaimheadfont@\enspace}}\proclaimheadfont@
   \ignorespaces##1\unskip\frills@{.\enspace}%
  \it\ignorespaces}%
 \nofrillscheck\proclaim}
\def\endproclaim{\revert@envir\endproclaim \par\rm
  \penaltyandskip@{55}\medskipamount}

\def\demoheadfont@{\smc}
\def\remarkheadfont@{\smc}

\def\enddemo{\par\revert@envir\enddemo \endremark\vskip\z@ plus 1\p@\relax}

\def\remark{\let\savedef@\remark \let\remark\relax
  \add@missing\endroster \add@missing\endproclaim
  \envir@stack\endremark
  \def\remark##1{\restoredef@\remark
    \penaltyandskip@\z@\smallskipamount
  {\def\usualspace{{\remarkheadfont@\enspace}}%
  \varindent@\remarkheadfont@\ignorespaces##1\unskip%
  \frills@{.\enspace}}\rm
  \ignorespaces}\nofrillscheck\remark}
\def\demo{\DN@{\ifx\next\nofrills
    \DN@####1####2{\remark####1{####2}\envir@stack\enddemo
      \ignorespaces}%
  \else
    \DN@####1{\remark{####1}\envir@stack\enddemo\ignorespaces}%
  \fi
  \next@}%
\FN@\next@}



\def\citefont@{\bf}

\let\Refs\relax
\outer\def\Refs{\add@missing\endroster \add@missing\endproclaim
 \let\savedef@\Refs \let\Refs\relax 
 \def\Refs##1{\restoredef@\Refs
   \if\notempty{##1}\penaltyandskip@{-200}\aboveheadskip
     \begingroup \raggedcenter@\refheadfont@
       \ignorespaces##1\endgraf\endgroup
     \penaltyandskip@\@M\belowheadskip
   \fi
   \begingroup\def\envir@end{\endRefs}\refsfont@\sfcode`\.\@m
   }%
 \nofrillscheck{\csname Refs\expandafter\endcsname
  \frills@{{References}}}}



\outer\def\enddocument{\par
  \add@missing\endRefs
  \add@missing\endroster \add@missing\endproclaim
  \add@missing\enddefinition
  \add@missing\enddemo \add@missing\endremark \add@missing\endexample
 \ifmonograph@ 
 \else
 \nobreak
 \thetranslator@
 \count@\z@ \loop\ifnum\count@<\addresscount@\advance\count@\@ne
 \csname address\number\count@\endcsname
 \csname email\number\count@\endcsname
 \repeat
\fi
 \vfill\supereject\end}

\catcode`\@=13

\pagewidth{30pc}
\pageheight{48pc}
\hoffset=1.8cm

\leftheadtext{ Chari and Pressley}
\rightheadtext{ Quantum Affine Algebras and their Representations}
\topmatter
\title Quantum Affine Algebras \\
and their Representations
\endtitle
\author Vyjayanthi Chari and Andrew Pressley\endauthor
\address Department of Mathematics, University of California, Riverside, CA
92521.\endaddress
\email chari\@ucrmath.ucr.edu\endemail
\address Department of Mathematics, King's College, London, WC2R
2LS.\endaddress
\email udah207\@bay.cc.kcl.ac.uk\endemail
\subjclass Primary 17B37, 81R50; Secondary 16W30, 82B23
\endsubjclass
\abstract We prove a highest weight classification of the finite-dimensional
irreducible representations of a quantum affine algebra, in the spirit of
Cartan's classification of the finite-dimensional irreducible representations
of complex simple Lie algebras in terms of dominant integral weights. We also
survey what is currently known about the structure of these representations.
\endabstract
\thanks The first author was partially supported by NSF Grant
\#9207701\endthanks
\endtopmatter

\document
\head 1. Introduction
\endhead
Around 1985, V.G. Drinfel'd and M. Jimbo showed, independently, how to
associate to any symmetrizable Kac--Moody algebra $\ung$ over $\Bbb C$ a family
$U_q(\ung)$ of Hopf algebras, depending on a parameter $q\in\Bbb C^{\times}$,
and reducing (essentially) to the classical universal enveloping algebra
$U(\ung)$ when $q=1$. The introduction of quantum groups has opened up a
fascinating new chapter in representation theory; in addition, quantum groups
have turned out to have surprising connections with several areas of
mathematics (algebraic groups in characteristic $p$, knot theory, $\ldots$ )
and physics (two--dimensional integrable systems, conformal field theories,
$\ldots$ ).

Many of the applications of quantum groups (such as those in knot theory, for
example) depend on the fact that, if $\ung$ is finite--dimensional and $q$ is
not a root of unity, one can associate to any finite--dimensional
representation $V$ of $\uqg$ an operator $R\in\text{End}\, (V\ot V)$ which
satisfies the quantum Yang--Baxter equation (QYBE):
$$R_{12}R_{13}R_{23} = R_{23}R_{13}R_{12}\tag1$$
(here, $R_{12}$ means $R\ot\text{id}\in\text{End}(V\ot V\ot V)$, etc.). In
fact, if $W$ is another finite--dimensional representation of $\uqg$, it turns
out that the tensor products $V\ot W$ and $W\ot V$ are isomorphic as
representations of $\uqg$, and further that there is a {\it canonical} choice
of isomorphism $I_{V,W}:V\ot W\to W\ot V$. If $V =W$ and $\sigma$ is the flip
map $V\ot V\to V\ot V$, the matrix $R =\sigma I$ satisfies (1).

In some situations, however, it is important to have a solution of the \lq QYBE
with spectral parameters\rq :
$$R_{12}(u,v)R_{13}(u,w)R_{23}(v,w) = R_{23}(v,w)R_{13}(u,w)R_{12}(u,v).\tag2$$
Here, $R(u,v)$ is a family of operators in $\text{End}\, (V\ot V)$, for some
finite--dimensional vector space $V$, depending on a pair of complex parameters
$u, v$. In many cases, possibly after making a change of variable $u\mapsto
f(u)$, $v\mapsto f(v)$, $R(u,v)$ becomes a function of $u-v$, which we write as
$R(u-v)$.

In  the theory of two--dimensional lattice models in statistical mechanics, for
example,
  $R(u)$ is a matrix whose entries are the \lq interaction\rq\  energies
of the atoms in the lattice, and $u$ is a parameter on which the properties of
the model depend, such as the values of external electric or magnetic fields.
{}From $R(u)$ one constructs the \lq transfer matrices\rq\
$$T(u) =R_{01}(u)R_{02}(u)\ldots R_{0N}(u)\in\text{End}\,(V\ot V^{\ot N})$$
(the first copy of $V$ in $V\ot V^{\ot N}$ is numbered $0$, the others
$1,\ldots ,N$) and from these the partition function
$$Z =\text{trace}_{V^{\ot N}}(\text{trace}_V(T)^N)$$
(we assume that the lattice is $N$ atoms wide in each direction and that
periodic boundary conditions are imposed). It is explained in [1], for example,
that the physical properties of the model may be deduced from $Z$. If $R(u)$ is
invertible and satisfies (2), it is easy to show that $\text{trace}_V(T(u))$
commutes with $\text{trace}_V(T(v))$ for all $u,v$: for this reason, such
models are called \lq integrable\rq .

One can hope to construct solutions of (2) whenever one has a Hopf algebra $A$
equipped with a family of automorphisms $\tau_u$.
For, if $V$ is a finite--dimensional (complex)
representation of $A$, pulling back $V$ by $\tau_u$ gives a 1-parameter family
of representations $V(u)$. Assume that, for all parameters $u$, $v$, $w$, and
for some representation $V$of $A$,

\roster
\item"(i)" $V(u)\ot V(v)$ is isomorphic to $V(v)\ot V(u)$, and

\item"(ii)" $V(u)\ot V(v)\ot V(w)$ is irreducible,
\endroster
and let $I(u,v): V(u)\ot V(v)\to V(v)\ot V(u)$ be an intertwiner (which, by
(i), is well--defined up to a scalar multiple). If $R =\sigma I$, equation (2)
is the condition for the equality of the two composites of intertwiners
\vskip6pt
{\eightpoint{$V(u)\ot V(v)\ot V(w)\to V(v)\ot V(u)\ot V(w)\to V(v)\ot V(w)\ot
V(u)\to
 V(w)\ot V(v)\ot V(u)$}}
\vskip6pt\noindent  and
\vskip6pt
{\eightpoint{$V(u)\ot V(v)\ot V(w)\to V(u)\ot V(w)\ot V(v)\to V(w)\ot V(u)\ot
V(v)\to V(w)\ot V(v)\ot V(u).
$}}
\vskip6pt\noindent Thus, condition (ii) guarantees that (2) is satisfied up to
a scalar multiple.

Let $\ungh$ be the (untwisted) affine Lie algebra associated to a
finite--dimensional complex simple Lie algebra $\ung$. Recall that $\ungh$ is a
central extension, with $1$--dimensional centre, of the Lie algebra
$\ung[t,t^{-1}]$ of Laurent polynomial maps $\Bbb{C^{\times}}\to\ung$, under
pointwise operations. There is an obvious multiplicative 1-parameter group of
automorphisms of $\ungh$, given by rescaling $t$, which fixes each element of
the centre. On the other hand, $\ungh$ is a symmetrizable Kac--Moody algebra,
so one can define the Hopf algebra $\uqgh$. We shall assume from now on that
$q$ is transcendental. It turns out that $\uqgh$ also has a multiplicative
1-parameter
group of automorphisms $\tau_u$, which reduce, in the limit $q\to 1$, to the
rescaling automorphisms of $\ungh$. According to Drinfel'd [11], property (i)
holds for generic values of $u$, $v$, and there exists a canonical choice of
isomorphism $I(u,v)$ such that $R(u,v)=\sigma I(u,v)$ satisfies (2). Moreover,
the multiplicative property of $\tau_u$ implies that $R(u,v)$ depends only on
$u/v$; reparametrizing by $u\mapsto e^u$, $v\mapsto e^v$, we get a solution of
(2) which depends only on $u-v$. Thus, it is of considerable interest to
describe the finite--dimensional irreducible representations of $\uqgh$.

The main result proved in this paper (Theorem 3.3) gives a parametrization of
the finite--dimensional irreducible representations of $\uqgh$ analogous to
Cartan's highest weight classification of the finite--dimensional
irreducible representations of $\ung$. The role of dominant integral weights in
the representation theory of $\ung$ is played for $\uqgh$ by the set of
rank($\ung$)--tuples $\bp$ of polynomials in one variable with constant
coefficient 1; let  $V(\bp)$ be the representation of $\uqgh$ associated to
$\bp$.

To construct explicit solutions of (2), one needs to understand the structure
of the representations $V(\bp)$. Now, there is a canonical embedding of Hopf
algebras $\uqg\hookrightarrow\uqgh$ which, in the limit $q\to 1$, becomes the
embedding $\ung\hookrightarrow\ungh$ given by regarding elelments of $\ung$ as
constant maps $\Bbb C^{\times}\to\ung$. Thus, representations of $\uqgh$ can be
regarded as representations of $\uqg$. Since finite-dimensional representations
of $\uqg$ are completely reducible,  a first step in understanding $V(\bp)$
would be to describe its decomposition under $\uqg$. We shall say that two
representations  of $\uqgh$ are {\it equivalent} if they are isomorphic as
representations of $\uqg$. Unfortunately, the problem of describing the
structure of $V(\bp)$ as a representation of $\uqg$ appears  to be intractable
for general $\bp$. However, it is still interesting to understand the
representations $V(\bp)$ of some special type.

Any $V(\bp)$ has a unique irreducible $\uqg$--subrepresentation of maximal
highest weight. Conversely, given a finite--dimensional irreducible
representation $V$ of $\uqg$, one can consider the representations $V(\bp)$ of
$\uqgh$ which have $V$ as their top $\uqg$--component -- $V(\bp)$ is then
called an {\it affinization} of $V$. (Thus,  every $V(\bp)$ is an affinization
of its top $\uqg$-component.)

Classically, every finite--dimensional representation $V$ of $\ung$ has an
affinization (in the obvious sense) which is irreducible under $\ung$. For,
there is an algebra homomorphism $ev_u:\ungh\to\ung$, for any $u\in\Bbb
C^{\times}$, which annihilates the centre of $\ungh$ and evaluates maps $\Bbb
C^{\times}\to \ung$ at $u$; note that $ev_u$ is the identity on $\ung$.
Pulling back $V$ by $ev_u$ gives a family of representations $V(u)$ of $\ungh$,
which are obviously isomorphic to $V$ as representations of $\ung$. In the
quantum case, however, there are simple examples of irreducible representations
of $\uqg$ which have no affinization that is irreducible under $\uqg$.
Thus, it is natural to look for the \lq smallest\rq\  affinization(s).

In [4], a natural partial ordering was defined  on the set of equivalence
classes of finite--dimensional representations  of $\uqg$. One can show that a
given irreducible representation $V$ of $\uqg$ has only finitely many
affinizations, up to equivalence, so it makes sense to look for the minimal
one(s). In Section 6, we give necessary and sufficient conditions on $\bp$ for
$V(\bp)$ to be a minimal affinization of its top $\uqg$--component, summarizing
results in [7], [9], [4], and [10]. We use these results to
describe the $\uqg$--structure of the minimal affinizations in some cases.

\head 2. Quantum affine algebras
\endhead
We begin by recalling the definition of the Hopf algebras $\uqg$.
Let $\ung$ be a finite--dimensional complex simple Lie algebra with Cartan
subalgebra $\unh$ and Cartan matrix $A=(a_{ij})_{i,j\in I}$. Fix coprime
positive integers  $(d_i)_{i\in I}$\/ such that $(d_ia_{ij})$\/ is symmetric.
Let $P=\Bbb Z^I$ and let
$P^+=\{\lambda\in P\mid \lambda(i)\ge 0\ \text{for all $i\in I$}$\}. For $i\in
I$, define $\lambda_i\in P^+$ by $\lambda_i(j) =\delta_{ij}$. Let $R$ (resp.
$R^+$) be the set of roots (resp. positive roots) of $\ung$. Let  $\alpha_i$
($i\in I$) be the simple roots and let $\theta$ be the highest root. Let $Q =
\op_{i\in I}\Bbb Z.\alpha_i\subset\unh ^*$\/ be the root lattice, and set $Q^+
=\sum_{i\in I}\Bbb N.\alpha_i$. Define a partial order $\ge$ on $P$ by
$\lambda\ge \mu$ iff $\lambda-\mu\in Q^+$. If $\eta =\sum_{i\in
I}m_i\alpha_i\in Q^+$, define $\text{height} (\eta )=\sum_{i\in I}m_i$.  Define
a non-degenerate symmetric bilinear form $(\ ,\ )$ on $\unh^*$ by
$(\alpha_i,\alpha_j)=d_ia_{ij}$,
and set $d_0=\frac12(\theta,\theta)$.

Let $q\in \Bbb C^{\times}$ be transcendental, and, for $r,n\in\Bbb N$, $n\ge
r$, define
$$\align [n]_q & =\frac{q^n -q^{-n}}{q -q^{-1}},\\
[n]_q! &=[n]_q[n-1]_q\ldots [2]_q[1]_q,\\
\left[{n\atop r}\right]_q &= \frac{[n]_q!}{[r]_q![n-r]_q!}.\endalign$$
Set $q_i=q^{d_i}$.

\proclaim{Proposition 2.1} There is a Hopf algebra $\uqg$ over $\Bbb C$ which
is generated as an algebra by elements $x_i^{{}\pm{}}$, $k_i^{{}\pm 1}$ ($i\in
I$), with the following defining relations:
$$\align
k_ik_i^{-1} = k_i^{-1}k_i &=1,\;\;  k_ik_j =k_jk_i,\\
k_ix_j^{{}\pm{}}k_i^{-1} &= q_i^{{}\pm a_{ij}}x_j^{{}\pm},\\
[x_i^+ , x_j^-] &= \delta_{ij}\frac{k_i - k_i^{-1}}{q_i -q_i^{-1}},\\
\sum_{r=0}^{1-a_{ij}}\left[{{1-a_{ij}}\atop r}\right]_{q_i}
(x_i^{{}\pm{}})^rx_j^{{}\pm{}}&(x_i^{{}\pm{}})^{1-a_{ij}-r} =0, \ \ \ \ i\ne
j.\endalign$$

The comultiplication $\Delta$, counit $\epsilon$, and antipode $S$ of $\uqg$
are given by
$$\align\Delta(x_i^+)&= x_i^+\ot k_i +1\ot x_i^+,\\
\Delta(x_i^-)&= x_i^-\ot 1 +k_i^{-1}\ot x_i^-,\\
\Delta(k_i^{{}\pm 1}) &= k_i^{{}\pm 1}\ot k_i^{{}\pm 1},\\
\epsilon(x_i^{{}\pm{}}) =0,\;\ & \epsilon(k_i^{{}\pm 1}) =1,\\
S(x_i^+) = -x_i^+k_i^{-1},\; S(x_i^-) &=- k_ix_i^-, \; S(k_i^{{}\pm 1})
=k_i^{{}\mp 1},\endalign$$
for all $i\in{I}$.
\endproclaim
The generators and relations in 2.1 serve, in fact, to define a Hopf algebra
$\uqg$ when $\ung$ is an arbitrary symmetrizable Kac--Moody algebra.
In particular, if $\ungh$ is the (untwisted) affine Lie algebra associated to
$\ung$, one can define the Hopf algebra $\uqgh$ as in 2.1, but replacing $I$ by
$\hat{I} = I\amalg\{0\}$ and $A$ by the extended Cartan matrix $\hat{A}
=(a_{ij})_{i,j\in\hat{I}}$ of $\ung$; we let  $q_0=q^{d_0}$.

Note that there is a canonical homomorphism $\uqg\to\uqgh$ such that
$x_i^{{}\pm{}}\mapsto x_i^{{}\pm{}}$, $k_i^{{}\pm 1}\mapsto k_i^{{}\pm 1}$ for
all $i\in I$. Thus, any representation of $\uqgh$ may be regarded as a
representation of $\uqg$.

Now $\ungh$ is better understood than an arbitrary infinite--dimensional
Kac--Moody Lie algebra because it has another realization as (a central
extension of)
a space of maps
 $\Bbb C^{\times}\to\ung$, as we mentioned in Section 1. In [12], Drinfel'd
stated (in a slightly different form) a realization of $\uqgh$ which, although
still in terms of generators and relations, more closely resembles the
description of $\ungh$ as a space of maps.
In the following form, the result was proved by Beck [2]:

\proclaim{Theorem 2.2}
Let  ${\Cal A}_q$ be the algebra with generators $x_{i,r}^{{}\pm{}}$ ($i\in I$,
$r\in\Bbb Z$), $k_i^{{}\pm 1}$ ($i\in I$), $h_{i,r}$ ($i\in I$, $r\in \Bbb
Z\backslash\{0\}$) and $c^{{}\pm{1/2}}$, and the following defining relations:
$$\align
c^{{}\pm{1/2}}\ &\text{are central,}\\
k_ik_i^{-1} = k_i^{-1}k_i =1,\;\; &c^{1/2}c^{-1/2} =c^{-1/2}c^{1/2} =1,\\
k_ik_j =k_jk_i,\;\; &k_ih_{j,r} =h_{j,r}k_i,\\
k_ix_{j,r}k_i^{-1} &= q_i^{{}\pm a_{ij}}x_{j,r}^{{}\pm{}},\\
[h_{i,r} , x_{j,s}^{{}\pm{}}] &= \pm\frac1r[ra_{ij}]_{q_i}c^{{}\mp
{|r|/2}}x_{j,r+s}^{{}\pm{}},\\
x_{i,r+1}^{{}\pm{}}x_{j,s}^{{}\pm{}} -q_i^{{}\pm
a_{ij}}x_{j,s}^{{}\pm{}}x_{i,r+1}^{{}\pm{}} &=q_i^{{}\pm
a_{ij}}x_{i,r}^{{}\pm{}}x_{j,s+1}^{{}\pm{}}
-x_{j,s+1}^{{}\pm{}}x_{i,r}^{{}\pm{}},\tag3\\
[x_{i,r}^+ , x_{j,s}^-]=\delta_{ij} & \frac{ c^{(r-s)/2}\phi_{i,r+s}^+ -
c^{-(r-s)/2} \phi_{i,r+s}^-}{q_i - q_i^{-1}},\endalign$$
\vskip6pt\noindent
{\rm{(4)}}\hskip0.5cm$\sum_{\pi\in\Sigma_m}\sum_{k=0}^m(-1)^k\left[{m\atop
k}\right]_{q_i} x_{i, r_{\pi(1)}}^{{}\pm{}}\ldots x_{i,r_{\pi(k)}}^{{}\pm{}}
x_{j,s}^{{}\pm{}}
 x_{i, r_{\pi(k+1)}}^{{}\pm{}}\ldots x_{i,r_{\pi(m)}}^{{}\pm{}} =0$,
\vskip6pt\noindent
if $i\ne j$, for all sequences of integers $r_1,\ldots, r_m$, where $m
=1-a_{ij}$, $\Sigma_m$ is the symmetric group on $m$ letters, and the
$\phi_{i,r}^{{}\pm{}}$ are determined by equating powers of $u$ in the formal
power series
$$\sum_{r=0}^{\infty}\phi_{i,\pm r}^{{}\pm{}}u^{{}\pm r} = k_i^{{}\pm 1}
exp\left(\pm(q_i-q_i^{-1})\sum_{s=1}^{\infty}h_{i,\pm s} u^{{}\pm s}\right).$$

If $\theta =\sum_{i\in I}m_i\alpha_i$, set $k_{\theta} = \prod_{i\in
I}k_i^{m_i}$. Suppose that the root vector $\overline{x}_{\theta}^+$ of $\ung$
corresponding to $\theta$ is expressed in terms of the simple root vectors
$\overline{x}_i^+$ ($i\in I$) of $\ung$ as
$$\overline{x}_{\theta}^+ = \lambda[\overline{x}_{i_1}^+, [\overline
x_{i_2}^+,\cdots ,[\overline x_{i_k}^+, \overline x_j^+]\cdots ]]$$
for some $\lambda\in\Bbb C^{\times}$. Define maps $w_i^{{}\pm{}}:\uqgh\to\uqgh$
by
$$w_i^{{}\pm{}}(a) = x_{i,0}^{{}\pm{}}a - k_i^{{}\pm 1}ak_i^{{}\mp
1}x_{i,0}^{{}\pm{}}.$$
Then, there is an  isomorphism $f:\uqgh\to\Cal A_q$  defined on generators by
$$\align
f(k_0) = k_{\theta}^{-1}, \ f(k_i) &= k_i, \ f(x_i^{{}\pm{}}) =
x_{i,0}^{{}\pm{}},  \ \ \ \ (i\in I),\\
f(x_0^+) &=\mu w_{i_1}^-\cdots w_{i_k}^-(x_{j,1}^-)k_{\theta}^{-1},\\
f(x_0^-) &=\lambda k_{\theta} w_{i_1}^+\cdots w_{i_k}^+(x_{j,-1}^+),\endalign
$$
where $\mu\in\Bbb C^{\times}$ is determined by the condition
$$[x_0^+, x_0^-] =\frac{k_0-k_0^{-1}}{q_0-q_0^{-1}}. $$
\endproclaim

Let $\hat U^{{}\pm{}}$ (resp. $\hat U^0$) be the subalgebra of  $\uqgh$
generated by the $x_{i,r}^{{}\pm{}}$ (resp. by the $\phi_{i,r}^{{}\pm{}}$) for
all $i\in I$, $r\in\Bbb Z$. Similarly, let $U^{{}\pm{}}$ (resp. $U^0$) be the
subalgebra of $\uqg$ generated by the $x_i^{{}\pm{}}$ (resp. by the $k_i^{{}\pm
1}$) for all $i\in I$. We have the following weak version of the
Poincar\'e--Birkhoff--Witt theorem:
\proclaim{ Proposition 2.3}
\roster
\item"(a)" $\uqg = U^-.U^0.U^+.$

\item"(b)" $\uqgh = \hat U^-.\hat U^0.\hat U^+.$
\endroster
\endproclaim
See [8] or [14] for details.

\head 3. Representation theory of $\uqg$ and $\uqgh$
\endhead

We begin by summarizing the relevant facts about the representation theory of
$\uqg$ (we continue to assume that $q$ is transcendental). For further
details, see
[8] or [14], for example.

Let $W$ be a representation of $\uqg$. One says that $\lambda\in P$ is a weight
of $W$ if the weight space
$$W_{\lambda} =\{w\in W| k_i.w =q^{\lambda(i)} w\}$$
is non--zero. We say that $W$ is of type 1 if
$$W =\bigoplus_{\mu\in P}W_{\mu}.$$
A non--zero vector $w\in W_{\lambda}$ is called a highest weight vector if
$x_i^+.w =0$ for all $i\in I$, and $W$ is called a highest weight
representation with highest weight $\lambda$ if $W =\uqg.w$ for some highest
weight vector $w\in W_{\lambda}$. Any highest weight representation is of type
1.

For any $\lambda\in P$, let $M(\lambda)$ be the quotient of $\uqg$ by the left
ideal generated by $\{x_i^+, k_i -q^{\lambda(i)}.1\}_{i\in I}$. Then,
$M(\lambda)$ is a highest weight representation of $\uqg$ with highest weight
$\lambda$, and it follows from 2.3(a) that $M(\lambda)_{\lambda}$ is
one--dimensional. The standard argument implies that $M(\lambda)$ has a unique
irreducible quotient $V(\lambda)$, and that every irreducible highest weight
representation with highest weight $\lambda$ is isomorphic to $V(\lambda)$.

For any $i\in I$, let $\sigma_i$ be the algebra automorphism of $\uqg$ such
that
$$\sigma_i(x_j^+) = (-1)^{\delta_{ij}}x_j^+,\, \, \sigma_i(k_j) =
(-1)^{\delta_{ij}}k_j, \, \, \sigma_i(x_j^-) = x_j^- \tag5$$
for all $j\in I$.

\proclaim{Proposition 3.1}
\roster
\item "(a)" Every finite--dimensional representation of $\uqg$ is completely
reducible.

\item "(b)" Every finite--dimensional irreducible representation of $\uqg$ can
be obtained from a type 1 representation by twisting with a product of the
automorphisms $\sigma_i$.

\item "(c)" Every finite--dimensional irreducible representation of $\uqg$ of
type 1 is  highest  weight.

\item"(d)" The representation $V(\lambda)$ is finite--dimensional iff
$\lambda\in P^+$.

\item "(e)" If $\lambda\in P^+$,  $V(\lambda)$  has the same character as the
irreducible representation of $\ung$ of the same highest weight.

\item "(f)" The multiplicity $m_\nu(V(\lambda)\ot V(\mu))$ of $V(\nu)$
in the tensor product $V(\lambda)\ot V(\mu)$, where $\lambda ,\mu, \nu\in P^+$,
is the same as in the tensor product of the irreducible representations of
$\ung$ of the same highest weight (this statement makes sense in view of parts
(a), (c) and (d)).
\endroster
 \endproclaim
We now turn to the representation theory of $\uqgh$.
A representation $V$ of $\uqgh$ is of type 1 if $c^{1/2}$ acts as the identity
on $V$, and if $V$ is of type 1 as a representation of $\uqg$. A vector $v\in
V$ is a highest weight vector if
$$x_{i,r}^+.v=0,\ \ \phi_{i,r}^{{}\pm{}}.v=\Phi_{i,r}^{{}\pm{}}v,\ \ \ c^{1/2}.
v =v,$$
for some complex numbers $\Phi_{i,r}^{{}\pm{}}$. A type 1 representation $V$ is
a highest weight representation if $V=\uqgh.v$, for some highest weight vector
$v$, and the pair of {\hbox{$(I\times\Bbb Z)$}}--tuples
$(\Phi_{i,r}^{{}\pm{}})_{i\in I,r\in\Bbb Z}$ is its highest weight.  (In [8],
highest weight representations of $\uqgh$ are called `pseudo-highest weight'.)

 Note that $\Phi_{i,r}^+=0$ (resp. $\Phi_{i,r}^-=0$) if $r<0$ (resp. if $r>0$),
and that $\Phi_{i,0}^+\Phi_{i,0}^-=1$.
Conversely, if ${\pmb\Phi} =(\Phi_{i,r}^{{}\pm{}})_{i\in I, r\in\Bbb Z}$ is  a
set of complex numbers satisfying these conditions, let $M(\pmb\Phi)$ be the
quotient of $\uqgh$ by the left ideal generated by $\{x_{i,r}^+,
\phi_{i,r}^{{}\pm{}} - \Phi_{i,r}^{{}\pm{}} .1\}_{i\in I,r\in\Bbb Z}\cup
\{c^{{}\pm 1/2} -1\}$. Then, $M(\pmb\Phi)$ is a highest weight representation
of $\uqgh$. It follows from 2.3(b) that, regarding $M(\pmb\Phi)$ as a
representation of $\uqg$, we have $\text{dim} (M(\pmb\Phi))_{\lambda} = 1$, and
hence that $M(\pmb\Phi)$ has a unique irreducible quotient (as a representation
of $\uqgh$), say $V(\pmb\Phi)$. Clearly, every irreducible highest weight
representation
of $\uqgh$ is isomorphic to some $V(\pmb\Phi)$.

Let $\sigma_i$ ($i\in I$) be the algebra automorphisms of $\uqgh$ defined by
the formulas in (5), but with the indices $i,j\in\hat
{I}$. Also, let $\sigma$ be the algebra automorphism of $\uqgh$ given, in terms
of the presentation 2.2, by
$$\align
\sigma(c^{1/2})=-c^{1/2},\ & \ \ \sigma(x_{i,r}^{{}\pm{}}) =
(-1)^rx_{i,r}^{{}\pm{}},\\
\sigma(k_i)=k_i,\ & \ \ \sigma(h_{i,r})=h_{i,r}.\endalign$$
\proclaim{Proposition 3.2} Let $V$ be a finite-dimensional irreducible
representation of $\uqgh$.
\roster
\item "(a)" $V$ can be obtained from a type 1 representation by twisting with a
product of some of the automorphisms $\sigma_i$ ($i\in \hat{I}$), $\sigma$.
\item"(b)" If $V$ is of type 1 (as a representation of $\uqgh$), then $V$ is
highest weight.
\endroster
\endproclaim

See Section 12.2 of [8] for the proof.

Thus, to classify the finite-dimensional irreducible representations of
$\uqgh$, we have only to determine for which $\pmb\Phi$ the representation
$V(\pmb\Phi)$ is finite-dimensional. The answer to this question is the main
result of this paper. If $\lambda\in P^+$, let $\calp^\lambda$ be the set of
all $I$-tuples $(P_i)_{i\in I}$ of polynomials $P_i\in\Bbb C[u]$, with constant
term 1, such that $\deg(P_i)=\lambda(i)$ for all $i\in I$. Set
$\calp=\cup_{\lambda\in P^+}\calp^\lambda$.
\proclaim{Theorem 3.3} Let $\pmb\Phi=(\Phi_{i,r})_{i\in I,r\in\Bbb Z}$ be a
pair of $(I\times\Bbb Z)$-tuples of complex numbers, as above. Then, the
irreducible representation $V(\pmb\Phi)$ of $\uqgh$ is finite-dimensional iff
there exists $\bp=(P_i)_{i\in I}\in\calp$ such that
$$\sum_{r=0}^\infty
\Phi_{i,r}^+u^r=q_i^{\deg(P_i)}\frac{P_i(q_i^{-2}u)}{P_i(u)}=
\sum_{r=0}^\infty \Phi_{i,-r}^-u^{-r},\tag6$$
in the sense that the left- and right-hand terms are the Laurent expansions of
the middle term about $0$ and $\infty$, respectively. \endproclaim

By abuse of notation, we denote the finite-dimensional irreducible
representation of $\uqgh$ associated to $\bp$ by $V(\bp)$, and say that $\bp$
is its highest weight.

The \lq only if\rq\  part of 3.3 is proved in [8], and we shall say no more
about it in this paper. The \lq if\rq\  part is proved in the next two
sections.

To conclude the present section, however, we describe the behaviour of the
$I$-tuples $\bp$ under tensor products. If $\bp=(P_i)_{i\in I}$,
$\bq=(Q_i)_{i\in I}\in\calp$, let $\bp\ot\bq\in\calp$ be the $I$--tuple
$(P_iQ_i)_{i\in I}$.
\proclaim{Proposition 3.4} Let $\bp,\bq\in\calp$, and let $v_\bp$ and $v_\bq$
be $\uqgh$-highest weight vectors in $V(\bp)$ and $V(\bq)$, respectively. Then,
in $V(\bp)\ot V(\bq)$, we have
$$x_{i,r}^+.(v_\bp\ot v_\bq)=0,\ \ \phi_{i,r}^{{}\pm{}}.(v_\bp\ot
v_\bq)=\Psi_{i,r}^{{}\pm{}}(v_\bp\ot v_\bq)$$
for all $i\in I$, $r\in\Bbb Z$, where the complex numbers
$\Psi_{i,r}^{{}\pm{}}$ are related to $\bp\ot\bq$ as the $\Phi_{i,r}^{{}\pm{}}$
are related to $\bp$ in (6).\endproclaim

See [8] for the proof. The following result is an immediate consequence:
\proclaim{Corollary 3.5} Let $\bp,\bq\in\calp$. Then, $V(\bp\ot\bq)$ is
isomorphic, as a representation of $\uqgh$, to a quotient of the
subrepresentation of $V(\bp)\ot V(\bq)$ generated by the tensor product of the
highest weight vectors in $V(\bp)$ and $V(\bq)$.\endproclaim

Since every polynomial is a product of linear polynomials, the last result
suggests that we define a representation $V(\bp)$ of $\uqgh$ to be {\it
fundamental} if, for some $i\in I$, $P_j=1$ if $j\ne i$ and $\deg(P_i)=1$.
Then, iterating 3.5, we obtain
\proclaim{Corollary 3.6} For any $\bp\in\calp$, the representation $V(\bp)$ of
$\uqgh$ is isomorphic to a subquotient of a tensor product of fundamental
representations.\endproclaim

This suggests a method of proving the \lq if\rq\  part of Theorem 3.3. For, in
view of 3.6, it clearly suffices to prove that the fundamental representations
of $\uqgh$ are all finite-dimensional. Since the fundamentals are the \lq
simplest\rq\  representations of $\uqgh$, it should be possible to describe
them \lq explicitly\rq, and, in particular, to prove that they are
finite-dimensional. We shall use this approach in the $sl_2$ case in the next
section, and, although we have no doubt that it can be carried through in the
general case, we shall use a different, more abstract, approach to complete the
proof of 3.3 in Section 5.

\head 4. Proof of the main theorem: $sl_2$ case
\endhead
It is easy to construct finite--dimensional representations of the classical
affine Lie algebra $\ungh$ thanks to the existence of the family of
homomorphisms $ev_a: \ungh\to\ung$ which annihilate the centre of $\ungh$ and
evaluate maps $\Bbb C^{\times}\to\ung$ at $a\in\Bbb C^{\times}$. If $V$ is a
representation of $\ung$, the pull--back of $V$ by $ev_a$ is a  representation
$V_a$ of $\ungh$. Jimbo [13] defined an analogue of $ev_a$ for
$U_q(\hat{sl}_2)$:
\proclaim{Proposition 4.1} There is a family of algebra homomorphisms
$ev_a: U_q(\hat{sl}_2)\to U_q(sl_2)$, defined for all $a\in\Bbb C^{\times}$,
such that $ev_a(c^{1/2}) =1$ and
$$ev_a(x_{1,r}^+) = q^{-r}a^{-r}k_1^rx_1^+,\ \ \ ev_a(x_{1,r}^-) =
q^{-r}a^{-r}x_1^-k_1^r,$$
for all $r\in\Bbb Z$.\endproclaim
See [6], Proposition 4.1, for the proof.
\vskip 12pt

{\it Remark.} Jimbo defined an analogue of $ev_a$ for $U_q(\hat{sl}_n)$, for
all $n\ge 2$ (strictly speaking, if $n >2$ Jimbo's homomorphism takes values in
an \lq enlargement\rq\ of $U_q(sl_n)$). If $\ung$ is not of type $A$, there is
no homomorphism $\uqgh\to\uqg$ which is the identity on $\uqg\subset\uqgh$ (see
[8]).
\vskip 12pt

If $V$ is a type 1 representation of $U_q(sl_2)$, its pull--back $V_a$ by
$ev_a$ is obviously a type 1 representation of $U_q(\hat{sl}_2)$; we call $V_a$
an {\it evaluation representation} of $U_q(\hat{sl}_2)$. Since $ev_a$ is the
identity on $U_q(sl_2)$, $V_a$ is isomorphic to $V$ as a representation of
$U_q(sl_2)$; in particular, $V_a$ is irreducible if $V$ is. The
finite-dimensional irreducible type 1 representations of $U_q(sl_2)$ are easy
to describe. We know that there is exactly one such representation $V(r)$ of
each dimension $r+1\ge 1$, since the same is true for $sl_2$. It is easy to
check that, if $\{v_0,v_1,\ldots ,v_r\}$ is a basis of $V(r)$, the formulas
$$k_1.v_k =q^{r-2k} v_k,\ \ x_1^+.v_k = [r-k+1]_qv_{k-1},\ \ x_1^-.v_k
=[k+1]_qv_{k+1}$$
define the required representation (we set $v_{-1} = v_{r+1} =0$).
Using the relations in 2.2, it follows that $v_0$ is a $\uqgh$--highest weight
vector of $V(r)_a$, and that
$$\align \phi_{1,k}^{{}\pm{}}.v_0& = a^{-k}q^{k(r-1)}(q^r-q^{-r})v_0\\
h_{1,k}. v_0&= q^{-k}a^{-k}\frac{[rk]_q}{k}v_0.\endalign$$
Using these formulas, one finds that $V(r)_a\cong V(P_{r,a})$, where
$$P_{r,a}(u) =\prod_{k=1}^r(1 -a^{-1}q^{r-2k+1}u).$$
The set
$\Sigma_{r,a} =\{aq^{-r+1}, aq^{- r+3},\ldots , aq^{r-1}\}$ of roots of
$P_{r,a}$ is called the $q$-{\it segment} of length $r$ and centre $a$.

At this point, it is easy to complete the proof of 3.3 in the $sl_2$ case. As
we noted at the end of Section 3, it suffices to prove that the fundamental
representations are finite-dimensional. But, since $P_{1,a}(u)=1-a^{-1}u$, it
follows that the fundamental representations of $U_q(\hat{sl}_2)$ are precisely
the $V(1)_a$, for arbitrary $a\in\Bbb C^\times$. In particular, they all have
dimension 2.

Before turning to the general case of 3.3, however, we shall describe the
structure of the representations $V(P)$ of $U_q(\hat{sl}_2)$ in more detail:
\proclaim{Proposition 4.2} Let $r_1,r_2,\ldots,r_k\in\Bbb N$,
$a_1,a_2,\ldots,a_k\in\Bbb C^\times$, $k\in\Bbb N$. Then, the tensor product
$V(r_1)_{a_1}\ot V(r_2)_{a_2}\ot\cdots\ot V(r_k)_{a_k}$ is reducible as a
representation of $U_q(\hat{sl}_2)$ iff at least one pair of $q$-segments
$\Sigma_{r_i,a_i}$, $\Sigma_{r_j,a_j}$, $1\le i,j\le k$, are in special
position, in the sense that their union is a $q$-segment which properly
contains them both.\endproclaim

This is proved in [6].

It is now easy to describe the representation $V(P)$, for any polynomial
$P\in\Bbb C[u]$ with constant coefficient 1. The roots of $P$ form a {\it
multiset}, i.e. a finite set of non-zero complex numbers (the roots of $P$),
with a positive integer attached to each element of the set (its multiplicity
as a root of $P$). It is not difficult to show that every multiset can be
written uniquely as a union of $q$-segments, no two of which are in special
position. (The union is in the sense of multisets: the multiplicity of a
complex number in a union of multisets is the sum of its multiplicities in each
of them.) We can thus write
$$\text{multiset of roots of $P$} =
\Sigma_{r_1,a_1}\cup\Sigma_{r_2,a_2}\cup\cdots\cup\Sigma_{r_k,a_k}$$
for some $r_1,r_2,\ldots,r_k\in\Bbb N$, $a_1,a_2,\ldots,a_k\in\Bbb C^\times$,
$k\in\Bbb N$, and where no pair $\Sigma_{r_i,a_i}$, $\Sigma_{r_j,a_j}$ is in
special position. By 3.5 and 4.2, there is an isomorphism of representations of
$U_q(\hat{sl}_2)$
$$V(r_1)_{a_1}\ot V(r_2)_{a_2}\ot\cdots\ot V(r_k)_{a_k}\cong
V(P_{r_1,a_1}P_{r_2,a_2}\ldots P_{r_k,a_k}).$$
But, the polynomial $P_{r_1,a_1}P_{r_2,a_2}\ldots P_{r_k,a_k}$ has the same
roots as $P$, with the same multiplicities, and hence is equal to $P$ (both
polynomials having constant coefficient 1). Thus,
$$V(P)\cong V(r_1)_{a_1}\ot V(r_2)_{a_2}\ot\cdots\ot V(r_k)_{a_k}.$$
We have proved
\proclaim{Theorem 4.3} Every finite-dimensional irreducible representation of
$U_q(\hat{sl}_2)$ of type 1 is isomorphic to a tensor product of evaluation
representations.\endproclaim

There is an amusing interpretation of $q$-segments in terms of \lq
$q$-derivatives\rq, which will allow us to give a kind of Weyl dimension
formula for $V(P)$. We recall that, if $P\in\Bbb C[u]$, its $q$-derivative is
$$(D_qP)(u)=\frac{P(q^2u)-P(u)}{q^2u-u}.$$
It is obvious that $D_qP$ is a polynomial in $u$ (and $q$), and that
$$\lim_{q\to 1}D_qP=\frac{dP}{du}.$$
The interpretation we have in mind is based on the following elementary result,
whose proof we leave to the reader.
\proclaim{Proposition 4.4} Let $P\in\Bbb C[u]$ have non-zero constant
coefficient, and let $\Sigma_P$ be its multiset of roots. Then, for each
integer $k\ge 2$, the number of $q$-segments of length $k$ in $\Sigma_P$ is
equal to the number of common roots of the polynomials
$P,D_qP,\ldots,D_q^{k-1}P$.\endproclaim

To clarify the meaning of 4.4, suppose that, in the canonical decomposition of
$\Sigma_P$ into a union of $q$-segments, no two of which are in special
position, there is one segment of length 2 and one of length 3. Then, the
number of $q$-segments of length 2 in $\Sigma_P$ is 3:
\vskip6pt\centerline{
$\circ\underbrace{\phantom{xxxxxxx}}\circ$
\hskip2cm
$\circ\underbrace{\phantom{xxxxxxx}}\circ\underbrace{\phantom{xxxxxxx}}\circ$
}
\vskip18pt\noindent
Of course, there is one $q$-segment of length 3 in $\Sigma_P$, and none of
length $>3$.

In general, suppose that, for each $k\ge 1$, there are $n_k$ $q$-segments of
length $k$ in the canonical decomposition of $\Sigma_P$. Then, there are $N_k$
$q$-segments of length $k$ altogether, where
$$\align
N_1&=n_1+2n_2+3n_3+\cdots+rn_r,\\
N_2&=n_2+2n_3+3n_4+\cdots+(r-1)n_r,\\
&\vdots\\
N_r&=n_r,\endalign$$
and $r=\deg(P)$. Hence,
$$n_k=N_k-2N_{k+1}+N_{k+2}$$
(we set $N_k=0$ if $k>r$). By 4.3, and the discussion preceding it, it is clear
that
$$\dim(V(P))=\prod_{k=1}^r(k+1)^{n_k}.$$
A little rearrangement now gives
\proclaim{Proposition 4.5} For any $P\in\Bbb C[u]$ with constant coefficient 1,
$$\dim(V(P))=2^{\deg(P)}\prod_{k=2}^{\deg(P)}\left(\frac{k^2-1}{k^2}\right)^{N_k},$$
where, for each integer $k\ge 2$, $N_k$ is the number of common roots of
$P,D_qP,\ldots$ $\ldots,D_q^{k-1}P$.\endproclaim

It would be interesting to find an analogue of this result for the dimensions
of the representations $V(\bp)$ of $\uqgh$, for arbitrary $\ung$.

\head 5. Proof of the main theorem: general case
\endhead

Let $\bp=(P_i)_{i\in I}\in\calp$ and let $v_\bp$ be a $\uqgh$-highest weight
vector in $V(\bp)$. Since $\phi_{i,0}^{{}\pm{}}=k_i^{{}\pm 1}$, it follows from
(6) that, if we define $\lambda\in P^+$ by $\lambda(i)=\deg(P_i)$, then
$k_i.v_\bp=q_i^{\lambda(i)}v_\bp$, so $v_\bp\in V(\bp)_\lambda$. By 2.3(b),
$V(\bp)_\lambda=\Bbb C v_\bp$ and
$$V(\bp)=\bigoplus_{\eta\in Q^+}V(\bp)_{\lambda-\eta}.\tag7$$
Thus, to prove the \lq if\rq\  part of 3.3, it is enough to prove the following
assertions:
\roster
\item"(a)" $V(\bp)_{\lambda-\eta}=0$ for all except finitely many $\eta\in
Q^+$.
\item"(b)" For all $\eta\in Q^+$, $\dim(V(\bp)_{\lambda-\eta})<\infty$.
\endroster
\vskip6pt PROOF OF (a). Let $0\ne v\in V(\bp)_\mu$, where $\mu=\lambda-\eta$,
$\eta\in Q^+$. Let $U_i$ be the subalgebra of $\uqgh$ generated by
$x_i^{{}\pm{}}$ and $k_i^{{}\pm 1}$ ($i\in I$), and let $V_i=U_i.v$. Note that
there is an obvious homomorphism of algebras (actually an isomorphism)
$U_{q_i}(sl_2)\to U_i$ which takes $x_1^{{}\pm{}}\mapsto x_i^{{}\pm{}}$,
$k_1^{{}\pm 1}\mapsto k_i^{{}\pm 1}$, so $V_i$ may be regarded as a
representation of $U_{q_i}(sl_2)$. We claim that, to prove (a), it suffices to
prove
\roster\item"(c)" If $0\ne v\in V(\bp)_\mu$ and $V_i=U_i.v$, then
$\dim(V_i)<\infty$.
\endroster

To see that (c) implies (a), note that, if $s_i$ is the $i$th fundamental
reflection in the Weyl group $W$ of $\ung$, the finite-dimensionality of $V_i$
implies that its set of weights is stable under the action of $s_i$ (this
follows from 3.1(e) and the analogous classical statement). Hence,
$V(\bp)_\mu\ne 0$ implies $V(\bp)_{s_i(\mu)}\ne 0$ for all $i\in I$. It follows
that, if $w\in W$ is arbitrary, then $V(\bp)_{w(\mu)}\ne 0$. Since one can
choose $w$ so that $w(\mu)\in P^+$, it follows that any $\mu\in P$ such that
$V(\bp)_\mu\ne 0$ belongs to the finite set
$$W.\{\nu\in P^+\mid \nu\le\lambda\}.$$
Thus, we are reduced to proving (c).

Now (c) is clearly a consequence of
\roster\item"(d)" If $V(\bp)_\mu\ne 0$, there exists $N>0$ such that
$V(\bp)_{\mu-r\alpha_i}=V(\bp)_{\mu+r\alpha_i}=0$ if $r>N$.
\endroster

Indeed, assuming (d), it is clear that $V_i$ is spanned by
$\{(x_i^{{}\pm{}})^r.v\mid 0\le r\le N\}$.

To prove (d), note that it is obvious that $V(\bp)_{\mu+r\alpha_i}=0$ for
$r>>0$, since $\mu+r\alpha_i\le\lambda$ only for finitely many $r>0$. We shall
prove, on the other hand, that $V(\bp)_{\mu-r\alpha_i}=0$ if $r>3h+\lambda(i)$,
where $h={\roman{height}}(\lambda-\mu)$. Indeed, this follows from
\roster
\item"(e)" For any $r>0$, $V(\bp)_{\mu-r\alpha_i}$ is spanned by vectors of the
form
$$X_1^-x_{i_1,k_1}^-X_2^-x_{i_2,k_2}^-\ldots
X_h^-x_{i_h,k_h}^-X_{h+1}^-.v_\bp,\tag8$$
where $\lambda-\mu=\alpha_{i_1}+\alpha_{i_2}+\cdots+\alpha_{i_h}$,
$k_1,k_2,\ldots,k_h\in\Bbb Z$ are arbitrary, and each $X_p^-$, $1\le p\le h+1$,
is a product of the form
$$X_p^-=x_{i,\ell_{1,p}}^-x_{i,\ell_{2,p}}^-\ldots x_{i,\ell_{r_p,p}}^-,$$
for some $\ell_{1,p},\ell_{2,p},\ldots,\ell_{r_p,p}\in\Bbb Z$ and
$r_1,r_2,\ldots,r_{h+1}\in\Bbb N$ such that
$$r_1+r_2+\cdots+r_{h+1}=r$$ and
$$r_1,r_2,\ldots,r_h\le 3.\tag9$$
\endroster

To see that (e) implies that $V(\bp)_{\mu-r\alpha_i}=0$ if $r>3h+\lambda(i)$,
let $\hat{U}_i$ be the subalgebra of $\uqgh$ generated by
$\{x_{i,k}^{{}\pm{}},\phi_{i,k}^{{}\pm{}}\}_{k\in\Bbb Z}$, and set
$\hat{V}_i=\hat{U}_i.v_{\bp}$. There is an obvious homomorphism of algebras
(actually an isomorphism) $U_{q_i}(\hat{sl}_2)\to \hat{U}_i$ which takes
$x_{1,k}^{{}\pm{}}\mapsto x_{i,k}^{{}\pm{}}$, $\phi_{1,k}^{{}\pm{}}\mapsto
\phi_{i,k}^{{}\pm{}}$, so $\hat{V}_i$ may be regarded as a representation of
$U_{q_i}(\hat{sl}_2)$. According to Lemma 2.3 in [9], $\hat{V}_i\cong V(P_i)$
as a representation of $U_{q_i}(\hat{sl}_2)$ (in particular, $\hat{V}_i$ is
irreducible). It follows from 4.2 that $(\hat{V}_i)_{\lambda-s\alpha_i}=0$ if
$s>\lambda(i)$. On the other hand, 2.3 implies that
$V(\bp)_{\lambda-s\alpha_i}=(\hat{V}_i)_{\lambda-s\alpha_i}$ for all $s\ge 0$.
Now, for any vector (8) satisfying the conditions in (e), we have $r_{h+1}\ge
r-3h>\lambda(i)$, so $X_{h+1}^-.v_{\bp}\in V(\bp)_{\lambda-r_{h+1}\alpha_i}=0$

. Thus, (e) implies that $V(\bp)_{\lambda-r\alpha_i}=0$ if $r>3h+\lambda(i)$.

To prove (e), note that it is obvious by 2.3(b) that $V(\bp)_{\mu-r\alpha_i}$
is spanned by vectors of the form (8) satisfying all the stated conditions
except possibly condition (9). Thus, it suffices to show that any vector $v$ of
the form (8) which does not satisfy (9) can be written as a linear combination
of vectors of the same form which do satisfy (9). We prove this by induction on
$h$.

If $h=0$, there is nothing to prove. Assume that $h\ge 1$. By repeated use of
relation (4) in 2.2, the product $X_1^-x_{i_1,k_1}^-$ can be expressed as a
linear combination of terms of the form $Y_1^-x_{i_1,k_1}^-\tilde{Y}_1^-$,
where $Y_1^-$ and $\tilde{Y}_1^-$ are of the same form as $X_1^-$, but where
$Y_1^-$ is a product of $\le 3$ generators $x_{i,\ell}^-$, and $\tilde{Y}_1^-$
is a product of $\ge \ell_{r_1}-3$ such generators. (If $\ung$ is simply-laced,
we can assume that $Y_1^-$ is a single $x_{i,\ell}^-$, and if $\ung$ is of type
B, C or F, that $Y_1^-$ is a product of $\le 2$ such generators.) So $v$ can be
expressed as a linear combination of vectors
$$Y_1^-x_{i_1,k_1}^-\tilde{Y}_1^-X_2^-x_{i_2,k_2}^-\ldots X_{h+1}^-.v_\bp.$$
By the induction hypothesis,
$$\tilde{Y}_1^-X_2^-x_{i_2,k_2}^-\ldots X_{h+1}^-.v_\bp$$
can be expressed as a linear combination of vectors
$$Y_2^-x_{i_2,k_2}^-Y_3^-x_{i_3,k_3}^-\ldots Y_{h+1}^-.v_\bp,$$
where each of $Y_2^-,Y_3^-,\ldots,Y_h^-$ is a product of $\le 3$
$x_{i,\ell}^-$'s, and $Y_{h+1}^-$ is a product of $\ge r-3-3(h-1)=r-3$
$x_{i,\ell}^-$'s. This completes the inductive step and proves (e).

The proof of (a) is now complete.

\vskip6pt PROOF OF (b). We proceed by induction on $h={\roman{height}}(\eta)$.
If $\eta=0$, there is nothing to prove. If $\eta=\alpha_i$, we have to show
that the vectors $x_{i,k}^-.v_\bp$ ($k\in\Bbb Z$) span a finite-dimensional
space. But this space is obviously contained in $\hat{U}_i.v_\bp$, and we have
already seen that $\hat{U}_i.v_\bp$ is finite-dimensional.

Assume now that $h\ge 2$, and that (b) has been proved for $\eta$'s of height
$<h$. The weight space $V(\bp)_{\lambda-\eta}$ is spanned, in view of 2.3(b),
by vectors of the form
$$x_{i_1,k_1}^-x_{i_2,k_2}^-\ldots x_{i_h,k_h}^-.v_\bp,\tag10$$
where $\eta=\alpha_{i_1}+\alpha_{i_2}+\cdots+\alpha_{i_h}$ and
$k_1,k_2,\ldots,k_h\in\Bbb Z$. It clearly suffices to prove that the vectors
(10) span a finite-dimensional space for each fixed choice of $i_1,\ldots,i_h$;
denote this space by $V_{i_1,\ldots,i_h}$. By the induction hypothesis,
there exists $M\in\Bbb N$ such that, for all $i\in\{i_1,i_2,\ldots,i_h\}$,
$V(\bp)_{\lambda-\eta+\alpha_i}$ is spanned by vectors of the form
$$x_{j_2,\ell_2}^-x_{j_3,\ell_3}^-\ldots x_{j_h,\ell_h}^-.v_\bp,\tag11$$
where $\alpha_{j_2}+\alpha_{j_3}+\cdots+\alpha_{j_h}=\eta-\alpha_i$ and
$|\ell_2|,|\ell_3|,\ldots,|\ell_h|\le M$. It suffices to prove that
$V_{i_1,\ldots,i_h}$ is contained in the space
$$W=\sum_{k_2=-M}^{M+1} x_{i_2,k_2}^-.V(\bp)_{\lambda-\eta+\alpha_{i_2}}
+x_{i_1,0}^-.V(\bp)_{\lambda-\eta+\alpha_{i_1}},\tag12$$
since $W$ is finite-dimensional by the induction hypothesis.

For this, we prove, by induction on $k_1$, that the vector (10) lies in $W$ for
every $k_2,\ldots,k_h$ (we assume that $k_1\ge 0$, the proof for $k_1\le 0$
being essentially the same). The case $k_1=0$ is obvious. For the inductive
step, note that we can assume that $|k_2|,|k_3|,\ldots,|k_h|\le M$. Using
relation (3) in 2.2, any vector (10) can be written as a linear combination of
the vectors
$$\align
x_{i_2,k_2}^-x_{i_1,k_1}^-x_{i_3,k_3}^-\ldots x_{i_h,k_h}^- & .v_\bp,\tag13\\
x_{i_2,k_2+1}^-x_{i_1,k_1-1}^-x_{i_3,k_3}^-\ldots x_{i_h,k_h}^- &
.v_\bp,\tag14\\
x_{i_1,k_1-1}^-x_{i_2,k_2+1}^-x_{i_3,k_3}^-\ldots x_{i_h,k_h}^- & .v_\bp.\tag15
\endalign$$
But, vectors of types (13) and (14) obviously belong to $W$, and those of type
(15) belong to $W$ by the induction hypothesis on $k_1$. This completes the
inductive step. (Note that, by the induction hypothesis again, the vector
$$x_{i_2,k_2+1}^-x_{i_3,k_3}^-\ldots x_{i_h,k_h}^- .v_\bp$$
can be written as a linear combination of vectors
$$x_{i_2',k_2'}^-x_{i_3',k_3'}^-\ldots x_{i_h',k_h'}^-.v_\bp,$$
where $\alpha_{i_2'}+\cdots+\alpha_{i_h'}=\alpha_{i_2}+\cdots+\alpha_{i_h}$ and
$|k_2'|,\ldots,|k_h'|\le M$.)

This completes the proof of (b), and hence that of Theorem 3.3.

\head 6. Minimal affinizations \endhead

We saw at the beginning of Section 5 that, if $\bp=(P_i)_{i\in I}\in\calp$, and
$\lambda\in P^+$ is defined by $\lambda(i)=\deg(P_i)$, then
$$V(\bp)=\bigoplus_{\eta\in Q^+}V(\bp)_{\lambda-\eta}\ \ \text{and}\ \
\dim(V(\bp)_\lambda)=1.\tag16$$
Since $V(\bp)$ is finite-dimensional, it is completely reducible as a
representations of $\uqg$, and in view of (16) we have
$$V(\bp)\cong V(\lambda)\oplus\bigoplus_{\mu\in P^+}V(\mu)^{\oplus m_\mu}$$
as a representation of $\uqg$, where the multiplicities $m_\mu\in\Bbb N$ are
zero unless $\mu<\lambda$. Thus, $V(\bp)$ gives a way of extending the action
of $\uqg$ on $V(\lambda)$ to an action of $\uqgh$, at the expense of enlarging
$V(\lambda)$ by the addition of representations of $\uqg$ of smaller highest
weight. For this reason, we call $V(\bp)$ an {\it affinization} of
$V(\lambda)$. We say that two affinizations are {\it equivalent} if and only if
they are isomorphic as representations of $\uqg$, and we denote by $[V(\bp)]$
the equivalence class of $V(\bp)$.

There is one situation in which affinizations are unique, up to equivalence:
\proclaim{Proposition 6.1} For any $i\in I$, $V(\lambda_i)$ has a unique
affinization, up to equivalence.\endproclaim
\demo{Proof} If $V(\bp)$ is an affinization of $V(\lambda_i)$, then $P_j=1$ if
$j\ne i$ and $P_i(u)=1-a^{-1}u$, for some $a\in\Bbb C^\times$ (i.e. $V(\bp)$ is
a fundamental representation of $\uqgh$). Denoting this $V(\bp)$ by
$V(\lambda_i,a)$, we have to prove that the equivalence class
$[V(\lambda_i,a)]$ is independent of $a$.

We make use of the family of (Hopf) algebra automorphisms $\tau_t$ ($t\in\Bbb
C^\times$) of $\uqgh$ defined by
$$\tau_t(x_{i,k}^{{}\pm{}})=t^kx_{i,k}^{{}\pm{}},\ \
\tau_t(\phi_{i,k}^{{}\pm{}})=t^k\phi_{i,k}^{{}\pm{}},\ \
\tau_t(c^{1/2})=c^{1/2}.$$
It is easy to see that, for any $\bq=(Q_i)_{i\in I}\in\calp$, the pull-back
$\tau_t^*(V(\bq))$ of $V(\bq)$ by $\tau_t$ is isomorphic as a representation of
$\uqgh$ to $V(\bq^t)$, where $\bq^t=(Q_i^t)$ and
$$Q_i^t(u)=Q_i(tu).$$
In particular, $\tau_a^*(V(\lambda_i,a))\cong V(\lambda_i,1)$. Since $\tau_a$
is the identity of $\uqg$, it follows that $[V(\lambda_i,a)]=[V(\lambda_i,1)]$.
\enddemo
\proclaim{Corollary 6.2} For any $\lambda\in P^+$, $V(\lambda)$ has, up to
equivalence, only finitely many affinizations.\endproclaim
\demo{Proof} By 3.6, any affinization $V(\bp)$ of $V(\lambda)$ is isomorphic as
a representation of $\uqgh$ to a subquotient of a tensor product
$$\bigotimes_{i\in I}\bigotimes_{j=1}^{\lambda(i)}V(\lambda_i,b_{j,i}),$$
for some $b_{j,i}\in\Bbb C^\times$ (the order of the factors is unimportant).
By 6.1, this tensor product is, up to $\uqg$-isomorphism, independent of the
$b_{j,i}$. It therefore has only finitely many subquotients, regarded as a
representation of $\uqg$. \enddemo
\vskip12pt
In general, a representation $V(\lambda)$ of $\uqg$ has many inequivalent
affinizations, and it is natural to ask if one can make a canonical choice
among them. To this end, the following partial order on the set of
affinizations was introduced in [4].
\proclaim{Proposition 6.3} Let $\lambda\in P^+$ and let $V(\bp)$ and $V(\bq)$
be affinizations of $V(\lambda)$. Then, we write $[V(\bp)]\preceq[V(\bq)]$ iff,
for all $\mu\in P^+$, either
\roster\item"(i)" $m_\mu(V(\bp))\le m_{\mu}(V(\bq))$, or
\item"(ii)" there exists $\nu>\mu$ with $m_\nu(V(\bp))<m_\nu(V(\bq))$.
\endroster
Then, $\preceq$ is a partial order on the set of equivalence classes of
affinizations of $V(\lambda)$.\endproclaim

An affinization $V(\bp)$ of $V(\lambda)$ is {\it minimal} if, whenever $V(\bq)$
is an affinization of $V(\lambda)$ and $[V(\bq)]\preceq[V(\bp)]$, we have
$[V(\bp)]=[V(\bq)]$. In view of 6.2, minimal affinizations certainly exist.

If $\ung=sl_2$, we explained in Section 4 that the homomorphisms
$ev_a:U_q(\hat{sl}_2)\to U_q(sl_2)$ enable one to extend the action of
$U_q(sl_2)$ on any representation $V(\lambda)$ to an action of
$U_q(\hat{sl}_2)$ {\it on the same space}. These evaluation representations
obviously provide the unique minimal affinization. We mentioned in Section 4
that there are analogues of the $ev_a$ when $\ung=sl_n$ for any $n\ge 2$, so
the minimal affinizations are also unique, and irreducible under $\uqg$, in
that case.

The following result, proved in [7], gives the defining polynomials of the
minimal affinizations in the type A case.
\proclaim{Theorem 6.4} Let $\ung=sl_{n+1}(\Bbb C)$ and let $\lambda\in P^+$.
Number the nodes of the Dynkin diagram of $\ung$ as in [3]. Then, $V(\lambda)$
has, up to equivalence, a unique minimal affinization. It is represented by
$V(\bp)$, where $\bp=(P_i)_{i\in I}\in\calp^\lambda$, iff, for all $i\in I$
such that $\lambda(i)>0$, the roots of $P_i$ form a $q$-segment with centre
$a_i\in\Bbb C^\times$ (say) and length $\lambda(i)$, where
\roster\item"(i)" either, for all $i<j$ such that $\lambda(i)>0$ and
$\lambda(j)>0$,
$$\frac{a_i}{a_j}=q^{\lambda(i)+2(\lambda(i+1)+\cdots+\lambda(j-1))+\lambda(j)+j-i},$$
\item"(ii)" or,  for all $i<j$ such that $\lambda(i)>0$ and $\lambda(j)>0$,
$$\frac{a_j}{a_i}=q^{\lambda(i)+2(\lambda(i+1)+\cdots+\lambda(j-1))+\lambda(j)+j-i}.$$
\endroster
\endproclaim

To state the corresponding results when $\ung$ is of type B, C or F, number the
nodes of the Dynkin diagram as in [3], and define, for any $\lambda\in P^+$,
complex numbers $c_i(\lambda)$ as follows:
$$c_i(\lambda)=\cases q^{d_i(\lambda(i)+\lambda(i+1)+1)} & \text{if
$a_{i+1\,i}a_{i\,i+1}=1$,}\\
q^{d_i\lambda(i)+d_{i+1}\lambda(i+1)+2d_{i+1}-1} & \text{if
$a_{i+1\,i}a_{i\,i+1}=1$.}\endcases$$

\proclaim{Theorem 6.5} Let $\ung$ be non-simply-laced, and let $\lambda\in
P^+$. Then, $V(\bp)$ is a minimal affinization of $V(\lambda)$ iff
$\bp\in\calp^\lambda$ satisfies the following conditions:
\roster
\item"(i)" For all $i\in I$, either $P_i=1$ or the roots of $P_i$ form a
$q_i$-segment of length $\lambda(i)$ and centre $a_i$ (say).
\item"(ii)" Either, for all $i<j$ such that $\lambda(i)>0$ and $\lambda(j)>0$,
we have
$$\frac{a_i}{a_j}=\prod_{s=i}^{j-1}c_s,$$
or,  for all $i<j$ such that $\lambda(i)>0$ and $\lambda(j)>0$, we have
$$\frac{a_i}{a_j}=q^{2d_j-2d_i}\prod_{s=i}^{j-1}c_s^{-1}.$$
\endroster
The minimal affinization of $V(\lambda)$ is unique, up to equivalence.
\endproclaim

See [10] for the proof. Note that, for any $r$, $I\backslash I_r$ defines a
type A subdiagram, so 6.4 gives the precise conditions under which
$V(\bp_{I\backslash I_r})$ is a minimal affinization.

Turning finally to the D and E cases, we introduce the following notation. If
$\emptyset\ne J\subseteq I$, and $\lambda\in P^+$, let $\lambda_J$ be the
restriction of $\lambda:I\to\Bbb Z$ to $J$. Also, if $\bp=(P_i)_{i\in
I}\in\calp$, let $\bp_J$ be the $J$-tuple $(P_j)_{j\in J}$.

Let $i_0$ be the unique node of the Dynkin diagram of $\ung$ which is linked to
three other nodes. Then,
$$I\backslash\{i_0\}=I_1\amalg I_2\amalg I_3,$$
where $I_1,I_2$ and $I_3$ define type A subdiagrams.
\proclaim{Theorem 6.6} Let $\ung$ be of type D or E, let $\lambda\in P^+$, and
assume that $\lambda(i_0)\ne 0$.

If $\lambda_{I_r}=0$ for some $r\in\{1,2,3\}$, then $V(\lambda)$ has a unique
minimal affinization, up to equivalence. It is represented by $V(\bp)$ iff
$V(\bp_{I\backslash I_r})$ is a minimal affinization of $V(\lambda_{I\backslash
I_r})$.

If $\lambda_{I_r}\ne 0$ for all $r\in\{1,2,3\}$, then $V(\lambda)$ has exactly
three minimal affinizations, up to equivalence. In fact, $V(\bp)$ is a minimal
affinization of $V(\lambda)$ iff there exist $r\ne s$ in $\{1,2,3\}$ such that
$V(\bp_{I\backslash I_r})$ and $V(\bp_{I\backslash I_s})$ are minimal
affinizations of $V(\lambda_{I\backslash I_r})$ and $V(\lambda_{I\backslash
I_s})$, respectively. \endproclaim

See [9] for the proof.
\vskip12pt{\it Remark.} The result of this theorem no longer holds if we drop
the assumption $\lambda(i_0)>0$. If $\ung$ is of type $\text{D}_4$, for
example, and $\lambda(i_0)=0$, the number of minimal affinizations of
$V(\lambda)$ increases with $\lambda$ (roughly speaking), and is generally
greater than three.
\vskip12pt To conclude our discussion of minimal affinizations, we consider
their structure as representations of $\uqg$. Except when $\ung$ is of type A,
when the minimal affinizations are irreducible under $\uqg$, this is not well
understood. We give two results.
\proclaim{Theorem 6.7} Let $\ung$ be of type $\text{B}_2$, let $\theta$ be the
highest root of $\ung$, and assume that $\alpha_2$ is the short simple root.
Let $\lambda\in P^+$ and let $V(\bp)$ be a minimal affinization of
$V(\lambda)$. Then, as representations of $\uqg$,
$$V(\bp)\cong \bigoplus_{r=0}^{[\frac12\lambda(2)]}V(\lambda-r\theta).$$
\endproclaim

See [5] for the proof.
Our final result gives the $\uqg$-structure of most of the fundamental
representations of $\uqgh$.
\proclaim{Theorem 6.8} Number the nodes of the Dynkin diagram of $\ung$ as in
[3].

\noindent(a) $V(\lambda_i,1)\cong V(\lambda_i)$ under any of the following
conditions:
\roster
\item"(i)" $\ung$ is of type A or C and $i$ is arbitrary;
\item"(ii)" $\ung$ is of type $\text{B}_n$ ($n\ge 2$) and $i=1$ or $n$;
\item"(iii)" $\ung$ is of type $\text{D}_n$ ($n\ge 4$) and $i=1, n-1$ or $n$.
\endroster

\noindent(b) If $\ung$ is of type $\text{B}_n$ or $\text{D}_{n+1}$ ($n\ge 3$)
and $1<i<n$,
$$V(\lambda_i,1)\cong\bigoplus_{j=0}^{[i/2]}V(\lambda_{i-2j}).$$

\noindent(c) If $\ung$ is of type $\text{E}_6$,
$$\align
V(\lambda_1,1)&\cong V(\lambda_1),\ \ V(\lambda_2,1)\cong
V(\lambda_2)\oplus\Bbb C,\\
V(\lambda_3,1)&\cong V(\lambda_3)\oplus V(\lambda_6),\\
V(\lambda_4,1)&\cong V(\lambda_4)\oplus V(\lambda_1+\lambda_6)\oplus
V(\lambda_2)\oplus V(\lambda_2)\oplus\Bbb C,\\
V(\lambda_5,1)&\cong V(\lambda_5)\oplus V(\lambda_1),\ \ V(\lambda_6,1)\cong
V(\lambda_6).\endalign$$
(Here and below, $\Bbb C$ denotes the 1-dimensional trivial representation.)

\noindent(d) If $\ung$ is of type $\text{E}_7$,
$$\align
V(\lambda_1,1)&\cong V(\lambda_1)\oplus\Bbb C,\ \ V(\lambda_2,1)\cong
V(\lambda_2)\oplus V(\lambda_7),\\
V(\lambda_3,1)&\cong V(\lambda_3)\oplus V(\lambda_6)\oplus V(\lambda_1)\oplus
V(\lambda_1)\oplus\Bbb C,\\
V(\lambda_6,1)&\cong V(\lambda_6)\oplus V(\lambda_1)\oplus\Bbb C,\ \
V(\lambda_7,1)\cong V(\lambda_7).\endalign$$

\noindent(e) If $\ung$ is of type $\text{E}_8$,
$$\align
V(\lambda_1,1)&\cong V(\lambda_1)\oplus V(\lambda_8)\oplus\Bbb C,\\
V(\lambda_7,1)&\cong V(\lambda_7)\oplus V(\lambda_1)\oplus V(\lambda_8)\oplus
V(\lambda_8)\oplus \Bbb C,\\
V(\lambda_8,1)&\cong V(\lambda_8)\oplus\Bbb C.\endalign$$

\noindent(f) If $\ung$ is of type $\text{F}_4$,
$$\align
V(\lambda_1,1)&\cong V(\lambda_1)\oplus\Bbb C,\\
V(\lambda_2,1)&\cong V(\lambda_2)\oplus V(2\lambda_4)\oplus V(\lambda_1)\oplus
V(\lambda_1)\oplus\Bbb C,\\
V(\lambda_3,1)&\cong V(\lambda_3)\oplus V(\lambda_1),\ \ V(\lambda_4,1)\cong
V(\lambda_4).\endalign$$

\noindent(g) If $\ung$ is of type $\text{G}_2$,
$$V(\lambda_1,1)\cong V(\lambda_1)\oplus\Bbb C,\ \ V(\lambda_2,1)\cong
V(\lambda_2).$$
\endproclaim

This can be proved using the techniques of [5].


\Refs

\ref\no 1
\by R. J. Baxter
\book Exactly solved models in statistical mechanics
\publ Academic Press \publaddr New York \yr 1982
\endref

\ref\no 2
\by J. Beck
\paper Braid group action and quantum affine algebras
\jour preprint, MIT \yr 1993
\endref

\ref\no 3
\by N. Bourbaki
\book Groupes et alg\`ebres de Lie, Chapitres 4, 5 et 6
\publ Hermann \publaddr Paris \yr 1968
\endref

\ref\no 4
\by V. Chari
\paper Minimal affinizations of representations of quantum groups: the rank 2
case
\jour preprint \yr 1994
\endref

\ref\no 5
\bysame
\paper Minimal affinizations of representations of quantum
groups:\,$\uqg$-structure
\jour preprint \yr 1994
\endref

\ref\no 6
\by V. Chari and A. N. Pressley
\paper Quantum affine algebras
\jour Commun. Math. Phys. \vol 142 \yr 1991 \pages 261--283
\endref

\ref\no 7
\bysame
\paper Small representations of quantum affine algebras
\jour Lett. Math. Phys. \vol 30 \yr 1994 \pages 131--145
\endref

\ref\no 8
\bysame
\book A Guide to Quantum Groups \publ Cambridge University Press
\publaddr Cambridge \yr 1994
\endref

\ref\no 9
\bysame
\paper Minimal affinizations of representations of quantum groups: the
simply-laced case
\jour preprint \yr 1994
\endref

\ref\no 10
\bysame
\paper Minimal affinizations of representations of quantum groups: the
non-simply-laced case
\jour preprint \yr 1994
\endref

\ref\no 11
\by V. G. Drinfel'd
\paper Quantum groups
\inbook Proceedings of the International Congress of Mathematicians, Berkeley,
1986
\publ American Mathematical Society \yr 1987 \pages 798--820
\endref

\ref\no 12
\bysame
\paper A new realization of Yangians and quantized affine algebras
\jour Soviet Math. Dokl. \vol 36 \yr 1988 \pages 212--216
\endref

\ref\no 13
\by M. Jimbo
\paper A $q$-analog of $U(gl(N+1))$, Hecke algebra and the Yang--Baxter
equation
\jour Lett. Math. Phys. \vol 11 \yr 1986 \pages 247--252
\endref

\ref\no 14
\by G. Lusztig
\book Introduction to Quantum Groups
\publ Birkh\"auser \publaddr Boston \yr 1993
\endref

\endRefs
\enddocument